\documentclass[]{spie}  

\usepackage{color}
\usepackage{graphicx}
\usepackage{subfigure}
\usepackage{bm}
\usepackage{amsmath}
\usepackage{amssymb}
\usepackage{cite}
\usepackage{stfloats}
\usepackage{bm}
\usepackage{mathrsfs}

\usepackage{subfigure}
\usepackage{amsmath}
\usepackage{amssymb}
\usepackage{stfloats}

\title{Optimal illumination pattern for transport-of-intensity quantitative phase microscopy}

\author[]{Jiaji Li$^{1,2,3,4}$}
\author[]{Qian Chen$^{1,2}$}
\author[]{Jiasong Sun$^{1,2,3}$}
\author[]{Jialin Zhang$^{1,2,3}$}
\author[]{Xiangpeng Pan$^{1,2,3}$}
\author[]{Chao Zuo$^{1,2,3,*}$}

\affil[]{$^1$School of Electronic and Optical Engineering, Nanjing University of Science and Technology, No. 200 Xiaolingwei Street, Nanjing, Jiangsu Province 210094, China\\
$^2$Jiangsu Key Laboratory of Spectral Imaging \& Intelligent Sense, Nanjing University of Science and Technology, Nanjing, Jiangsu Province 210094, China\\
$^3$Smart Computational Imaging Laboratory (SCILab), Nanjing University of Science and Technology, Nanjing, Jiangsu Province 210094, China\\
$^4$jiajili@njust.edu.cn\\}

\authorinfo{Further author information: (Send correspondence to C. Zuo)\\E-mail: surpasszuo@163.com; zuochao@njust.edu.cn}

\begin{document}
\maketitle

\begin{abstract}
The transport-of-intensity equation (TIE) is a well-established non-interferometric phase retrieval approach, which enables quantitative phase imaging (QPI) of transparent sample simply by measuring the intensities at multiple axially displaced planes. Nevertheless, it still suffers from two fundamentally limitations. First, it is quite susceptible to low-frequency errors (such as ``cloudy'' artifacts), which results from the poor contrast of the phase transfer function (PTF) near the zero frequency. Second, the reconstructed phase tends to blur under spatially low-coherent illumination, especially when the defocus distance is beyond the near Fresnel region. Recent studies have shown that the shape of the illumination aperture has a significant impact on the resolution and phase reconstruction quality, and by simply replacing the conventional circular illumination aperture with an annular one, these two limitations can be addressed, or at least significantly alleviated. However, the annular aperture was previously empirically designed based on intuitive criteria related to the shape of PTF, which does not guarantee optimality. In this work, we optimize the illumination pattern to maximize TIE's performance based on a combined quantitative criterion for evaluating the ``goodness'' of an aperture. In order to make the size of the solution search space tractable, we restrict our attention to binary coded axis-symmetric illumination patterns only, which are easier to implement and can generate isotropic TIE PTFs. We test the obtained optimal illumination by imaging both a phase resolution target and HeLa cells based on a small-pitch LED array, suggesting superior performance over other suboptimal patterns in terms of both signal-to-noise ratio (SNR) and spatial resolution.
\end{abstract}

\section{Introduction}
The aim of phase imaging is to visualize or measure the phase of transparent object, such as the optical elements and biological samples, but the intensity images of transparent sample generally do not contain any phase information at the in-focus plane. Zernike phase contrast (PhC) microscopy \cite{PhC} and differential interference contrast (DIC) microscopy \cite{DIC} are used to enhance the phase contrast qualitatively in unstained, transparent samples by using the phase-shift objective and Nomarski prism. These two optical microscopy techniques convert the phase into intensity with the bright diffraction halo and bas-relief effect, respectively. However, the two visualization methods only provide the qualitative phase images of sample, and it is difficult to process the phase image for quantitative data interpretation like the cell counting and dry mass measurement.

Quantitative phase imaging (QPI) is an effective approach to recover the phase of unlabeled biological samples without phototoxicity or photobleaching as in fluorescence microscopy \cite{Flu}. Conventional off-axis digital holographic microscopy (DHM) \cite{DMH1,DMH2,DMH3} measures the total phase delay of sample by the spatial modulation of heterogeneous refractive index within the sample. And some other interference approaches utilizing the common path geometries and white light source have been presented to improve imaging resolution and quality of quantitative phase result \cite{GP1,GP2,GP3} as well. Moreover, the non-interferometric, deterministic phase retrieval algorithms existing for quantitative phase measurement have also been proposed in the several decades, just like the transport of intensity equation (TIE) \cite{TIE1,TIE2,TIE3,TIE4,TIE5,TIE6} and differential phase contrast (DPC) \cite{DPC_Colin}. These types of methods linearize the image formation process of phase object, and the QPI can be achieved by a direct inversion using the phase transfer function (PTF) \cite{Inv_QPI1,Inv_QPI2} of TIE or DPC.  The illumination source has also been extended to the programmable LED array illumination for two-dimensional (2D) and three-dimensional (3D) quantitative phase reconstruction results based on TIE and DPC algorithms \cite{TIE_LED_Lensless,TIE_LED_AI,DPC_LED,DPC_LED_3D}. Besides, the QPI can be realized by the Fourier ptychographic microscopy (FPM) based on an iterative recovery process as well in the recent work \cite{PFM_Phase1,PFM_Phase2,PFM_Phase3}.

DPC generates the complex optical transfer function for phase contrast through the asymmetric illumination, but the intensity images from multi directions are needed for the combination of PTF due to anisotropic PTF coverage in the whole Fourier spectrum \cite{DPC_LED}. As for the imaging system using the axis-symmetric source along optical axis, the defocusing of optical system along the $z$ axis is an another convenient way to introduce tunable complex PTF \cite{Inv_QPI1}. TIE is a typical approach for phase retrieval based on axial defocus, and the quantitative phase can be recovered by TIE only using object field intensities at multiple axially displaced planes \cite{TIE1}. But the transfer function of conventional TIE is a special case of PTF under weak defocus assumptions and coherent illumination, and the slow rising of PTF amplitude of TIE at low frequency usually leads the decrease of background contrast of phase image and the superimposition of cloud-like artifacts on the reconstructed phase \cite{TIE_Noise}. Thus, many previous works are concentrated on the noise alleviation of phase measurement through multi axial intensity images within the framework of TIE \cite{TIE_HighOrder,TIE_SGDF,TIE_MF}. Until these approaches have been further extended to account for partial coherence explicitly and the phase recovery based on TIE is well adapted to partially coherent illumination \cite{TIE_PC1,TIE_PC2,TIE_PC3,TIE_PC4,Partial_decomp}, the inverse Laplacian of TIE is replaced by the inverse PTF of partially coherent illumination under different defocus distances. Recent studies have shown that the shape of the illumination aperture has a significant impact on the resolution and phase reconstruction quality by simply replacing the conventional circular illumination aperture with an annular one \cite{TIE_LED_AI,AI_TIE,AI_ODT}, the imaging resolution and phase reconstruction quality can be addressed, or at least significantly alleviated. Although the annular aperture was empirically designed and have been applied to QPI, the optimality of this illumination pattern based on a combined quantitative criterion has not been proved yet.

In this work, we optimize the illumination pattern to maximize TIE's performance based on a combined quantitative criterion for evaluating the ``goodness'' of an aperture and implement this optimal illumination pattern to the LED source on a high density LED array in binary coding in conventional bright-field microscope. Firstly, the incoherent illumination source is divided into a lot of discrete annular illumination patterns with the same center, and these discrete annular patterns are encoded as binary code for the representation of arbitrary combination of illumination pattern. Then, the PTF of all combinations of binary illumination patterns under weak defocusing condition can be calculated, and the performance of these transfer functions is evaluated by three criteria: cutoff frequency, zero crossings number, and means of absolute value of transfer function. Finally, the PTF with the best imaging performance, which corresponds the maximum cutoff frequency, the minimum number of zero crossings, and the biggest mean value, is chosen as the optimal one among all combinations of illumination patterns.

Although too many works have been proposed for QPI based on the coherent and partially coherent illuminations (including circular and annular patterns), the novelty of this work is to estimate the quality of PTF quantitatively through numerical analysis instead of intuitive criteria related to the shape of PTF and guarantee the optimality of proposed illumination pattern. Additionally, the annular illumination pattern, whose outer radius equals objective pupil and  annular width is as small as possible, is identified as the optimal one for QPI. In contrast to the frequency filtering algorithm utilizing multi distance intensity images, only single defocus distance is used in this work to achieve the maximum frequency coverage of PTF in amplitude and cutoff frequency with optimal illumination pattern. Moreover, we test the obtained optimal illumination by imaging both a phase resolution target and HeLa cells based on a small-pitch LED array, suggesting superior performance over other suboptimal patterns in terms of both signal-to-noise ratio (SNR) and spatial resolution. And the theoretical analysis and experimental results of control sample and biological sample validate the optimality and success of this illumination pattern. Due to experimental simplicity, the proposed method is especially suitable for application to unstained phase objects, and may promote a more widespread adoption of QPI using optimal illumination source in the biomedical community.

\section{Principle}
\subsection{PTF for arbitrary isotropic illumination pattern}

Assuming that there is a phase object in bright-field microscope, and the function of object complex amplitude $t\left( {\bf{r}} \right)$ is described as $a\left( {\bf{r}} \right)\exp \left[ {i\phi \left( {\bf{r}} \right)} \right]$, where $a\left( {\bf{r}} \right)$ is the amplitude which characterizes the sample's absorption (equals 1 for pure phase object), and ${\phi \left( {\bf{r}} \right)}$ is phase, with $\bf{r}$ being the spatial coordinates. The image formation in partially coherent illumination system can be illustrated by the transmission cross coefficient (TCC) model \cite{DPC_LED,Hopkins,AI_TIE}, but the measured intensity is still nonlinear in the sample's absorption or phase. In order to simplify this problem, we adopt a weak object approximation, and the first-order Taylor expansion of sample's complex function can be expressed as:
\begin{equation}\label{Eq1}
t\left( {\bf{r}} \right) \equiv a\left( {\bf{r}} \right)\exp \left[ {i\phi \left( {\bf{r}} \right)} \right] \approx a\left( {\bf{r}} \right){\left[ {1 + i\phi \left( {\bf{r}} \right)} \right]^{a\left( {\bf{r}} \right) = {a_0} + \Delta a\left( {\bf{r}} \right)}} \approx {a_0} + \Delta a\left( {\bf{r}} \right) + i{a_0}\phi \left( {\bf{r}} \right)
\end{equation}
where $a_0$ is the mean value of amplitude of $a\left( {\bf{r}} \right)$. By doing so, the intensity image is the linear superposition of absorption and phase contribution \cite{Streibl,Nugent}, and the the cross terms from absorption and phase are neglected due to the weak scattered light of a weak phase object. The formula of intensity can be approximated as:
\begin{equation}\label{Eq2}
I\left( {\bf{r}} \right) = t\left( {\bf{r}} \right){t^ * }\left( {{\bf{r'}}} \right) \approx a_0^2{\rm{ + }}{a_0}\left[ {\Delta a\left( {\bf{r}} \right) + \Delta a\left( {{\bf{r'}}} \right)} \right]{\rm{ + }}i{a_0}\left[ {\phi \left( {\bf{r}} \right) + \phi \left( {{\bf{r'}}} \right)} \right]
\end{equation}
where $a_0^2$ is the background intensity, ${a_0}\left[ {\Delta a\left( {\bf{r}} \right) + \Delta a\left( {{\bf{r'}}} \right)} \right]$ is the absorption contrast, and ${a_0}\left[ {\phi \left( {\bf{r}} \right) + \phi \left( {{\bf{r'}}} \right)} \right]$ is the phase contrast, respectively. Further, we implement the Fourier transform to the intensity image, and the resulting spectrum of intensity is:
\begin{equation}\label{Eq3}
\widetilde I\left( {\bf{u}} \right) = B\delta \left( {\bf{u}} \right) + \widetilde A\left( {\bf{u}} \right){H_A}\left( {\bf{u}} \right) + \widetilde P\left( {\bf{u}} \right){H_P}\left( {\bf{u}} \right)
\end{equation}
where $\bf{u}$ represents the variable in Fourier polar coordinate, $B$ is the zero-frequency amplitude of intensity spectrum, $\widetilde P\left( {\bf{u}} \right)$, $\widetilde A\left( {\bf{u}} \right)$, ${H_P}\left( {\bf{u}} \right)$ and ${H_A}\left( {\bf{u}} \right)$ are the phase spectrum, absorption spectrum, PTF, and amplitude transfer function (ATF), accordingly. While the illumination is the oblique axis-symmetric source, the expression of weak object PTF has been derived in previous work \cite{TIE_LED_AI}:
\begin{equation}\label{Eq4}
\begin{aligned}
{{H}_{p}}{{\left( \mathbf{u} \right)}_{obl}}\text{=}& \frac{1}{2} \left| P\left( \mathbf{u}-{{\bm{\rho }}_{{s}}} \right) \right|\sin \left[ kz\left( \sqrt{1-{{\lambda }^{2}}{{\left| \mathbf{u}-{{\bm{\rho }}_{{s}}} \right|}^{2}}}-\sqrt{1-{{\lambda }^{2}}{{\left| {{\bm{\rho }}_{{s}}} \right|}^{2}}} \right) \right] \\
& + \frac{1}{2}\left| P\left( \mathbf{u}+{{\bm{\rho }}_{{s}}} \right) \right|\sin \left[ kz\left( \sqrt{1-{{\lambda }^{2}}{{\left| \mathbf{u}+{{\bm{\rho }}_{{s}}} \right|}^{2}}}-\sqrt{1-{{\lambda }^{2}}{{\left| {{\bm{\rho }}_{{s}}} \right|}^{2}}} \right) \right]
\end{aligned}
\end{equation}
where $z$ is the defocus distance along the optical axis, $k$ is the wave number, ${P\left( {\bf{u}} \right)}$ is objective pupil function, and ${{\bm{\rho }}_{{s}}}$ is the spatially normalized frequency of source. For this situation, the distribution of illumination source and objective pupil are defined as:
\begin{equation}\label{Eq5}
S\left( {\bf{u}} \right) = \delta \left( {{\bf{u}} - {{\bf{\rho }}_{{s}}}} \right){\rm{ + }}\delta \left( {{\bf{u}}{\rm{ + }}{{\bf{\rho }}_{{s}}}} \right)
\end{equation}
and
\begin{equation}\label{Eq6}
\left| P\left( \mathbf{u} \right) \right|=
\left\{
\begin{aligned}
& 1,\quad \text{if } \left| {\bf{u}} \right| \le {\rho _P} \\
& 0, \quad \text{if }\left| {\bf{u}} \right| > {\rho _P}
\end{aligned}
\right.
\end{equation}
where ${{\bf{\rho }}_{{s}}}$ and ${{\bf{\rho }}_{{p}}}$ are the spatially normalized frequency radius of source pupil and objective pupil in the Fourier space. Additionally, 2D images and line profiles of PTF for different types axis-symmetric sources under weak defocusing conditions are illustrated in the Fig. 1 in \cite{TIE_LED_AI}.

\begin{figure}[!t]
    \centering
    \includegraphics[width=16cm]{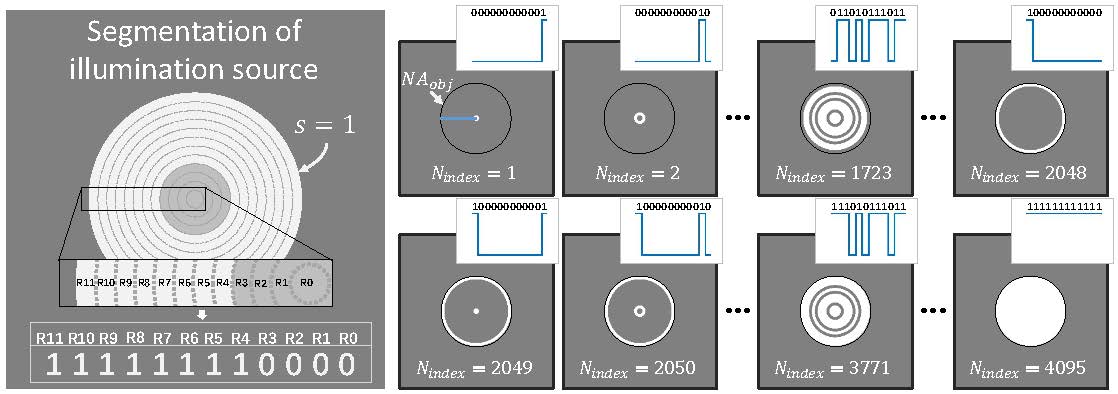}
    \caption{Segmentation of illumination source and labeling of all combinatorial illumination patterns. The incoherent illumination source is evenly divided into a lot of discrete annular patterns with same center point, and these annuli are labeled from the LSB $R0$ to the MSB $R11$ with 12-bit. The radius cut-profile of each illumination pattern is illustrated as a PWM wave or binary number, and the corresponding illumination pattern can be numbered from 1 to 4095 in decimal.}
    \label{Fig1}
\end{figure}

Due to the anisotropy of illumination source in Eq. (\ref{Eq5}), the maximum coverage of transfer function in Eq. (\ref{Eq4}) is only along the oblique direction of two delta points. While the annular source can provide the same coverage of PTF along all directions, and the PTF of arbitrary isotropic illumination pattern (including annular aperture and circular aperture) can be calculated by the incoherent superposition of all intensities of the coherent partial images arising from all discrete light source points on the source plane. For an incoherent K\"ohler illumination pattern, it can be described by a coherent decomposition using the theory of coherent mode decomposition \cite{Partial_decomp} with Eq. (\ref{Eq5}). The annular illuminating aperture contains a collection of coherent point sources, and the expression of a certain annular source can be written as following equation:
\begin{equation}\label{Eq7}
S({\bf{u}}) = \sum\limits_{i = 0}^n {\delta ({{\bf{u}}}- {{\bf{u}}_i} )}, \quad \left| {{{\bf{\rho }}_1}} \right| \le \left| {{{\bf{u}}_i}} \right| \le \left| {{{\bf{\rho }}_2}} \right|
\end{equation}
where $n$ is the number of all discrete light points satisfying $\left| {{{\bf{\rho }}_1}} \right| \le \left| {{{\bf{u}}_i}} \right| \le \left| {{{\bf{\rho }}_2}} \right|$ on the source plane, ${{{\bf{\rho }}_1}}$ and ${{{\bf{\rho }}_2}}$ are the normalized frequency of inner and outer radius of the annular source respectively. The value of ${{{\bf{\rho }}_1}}$ is always smaller than the value of ${{{\bf{\rho }}_2}}$, and the values of both ${{{\bf{\rho }}_1}}$ and ${{{\bf{\rho }}_2}}$ are smaller than the the maximum coherence parameter 1, where the coherence parameter $s$ is defined as the ratio of illumination NA to objective NA ($N{A_{ill}}/N{A_{obj}}$).

In order to find the optimal illumination pattern among all arbitrary isotropic illumination sources, the incoherent illumination source ($s = 1$) is evenly divided into a lot of discrete annular illumination patterns with same center point. And these discrete annular patterns with different inner and outer radius are encoded as a binary number with 12-bit, as shown in Fig. \ref{Fig1}. These rings on the source plane are labeled from the least significant bit (LSB) $R0$ to the most significant bit (MSB) $R11$, and the arbitrary combination of annular illumination pattern can be numbered as an unique binary number or a decimal number. The maximum number of binary illumination pattern is 4095 corresponding ${2^N} - 1$, and the maximum normalized frequency of combinatorial illumination pattern is 1 which corresponds to the objective NA, where $N$ is the bit-depth of segmentation of illumination source. Besides, the radius cut-profile of each illumination pattern is plot as the pulse width modulation (PWM) wave and the corresponding binary number is the index of illumination pattern from 1 to 4095 in decimal in Fig. \ref{Fig1}. The segmentation of illumination source and labeling of each combinatorial pattern makes it is easier to index arbitrary illumination pattern and calculate the PTF of corresponding source pattern.

\begin{figure}[!t]
    \centering
    \includegraphics[width=16cm]{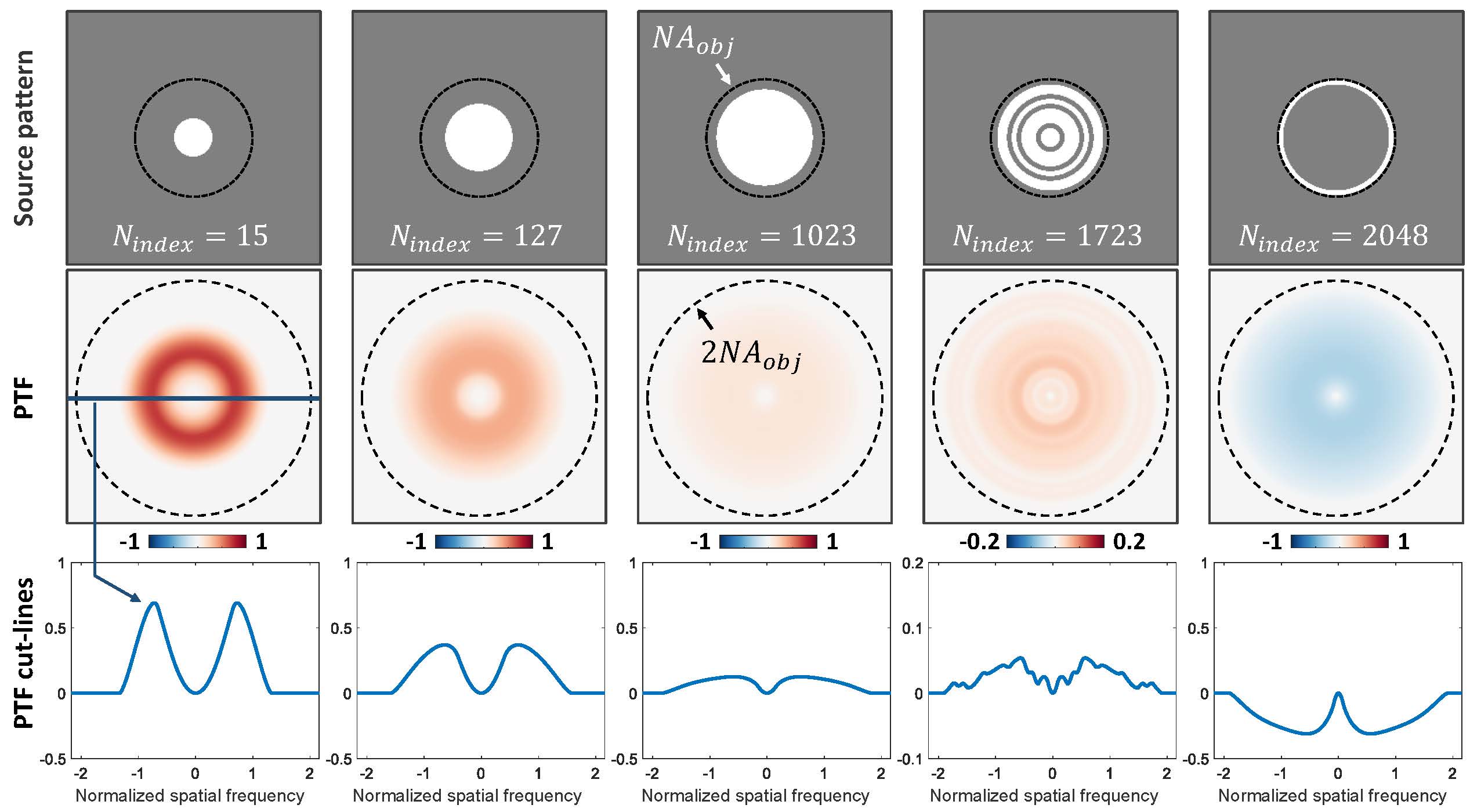}
    \caption{2D images and line profiles of PTF of five selected illumination patterns  under weak defocusing assumptions.}
    \label{Fig2}
\end{figure}

By invoking the expression of weak object PTF for oblique situation, the transfer functions of arbitrary illumination pattern can be calculated by Eq. (\ref{Eq4}) for a determined binary index illumination pattern further. Figure \ref{Fig2} shows the 2D PTF images and corresponding transfer function cut-lines of five selected illumination patterns under weak defocusing assumptions (0.5 $\mu$m). It can be seen that the cutoff frequency of PTF of circular source pattern is extended but the amplitude of transfer function response becomes weak from $N_{index} = 15$ to $N_{index} = 1023$. This phenomenon is consistent with previous conclusions in partially coherent illumination, and it is also a trade off between image contrast and resolution in the traditional bright-field microscope \cite{Streibl}. While the combinatorial illumination pattern is separated, the curve of transfer function becomes oscillating and uneven due to the offsetting and superposition of transfer function of inner and outer annular source with each other for pattern $N_{index} = 1723$. For the last column of Fig. \ref{Fig2} is the biggest single annular source and the transfer function is plotted as well. Not only the curve of PTF of illumination patten $N_{index} = 2048$ has relatively large cutoff frequency value, but also the curve is even in the whole passband.

\subsection{Criteria of PTF's quality}

Although the PTFs of all illumination pattern combinations are available, the quality of these transfer functions is hard to evaluate. For a signal system, the ideal transfer function of system filter is a rectangular function (amplitude of response equals 1) in the whole passband. Thus, the shape of the best PTF should be as close as possible to the shape of rectangular function, and three criteria are given for the analysis of PTF's quality as following: 1) The system cutoff frequency or imaging resolution can be extended to the maximum theoretical value. That is to say the achievable spatial frequency bandwidth of imaging system should be large. 2) The transfer function has fewer zero crossings. For the ideal situation, the curve of transfer function should be completely in the first quadrant or fourth quadrant, and there are no zero crossings excepting zero frequency point. 3) The mean of absolute value of transfer function or the area enclosed by the curve and coordinate axis should be large. Only in this case the transfer efficiency of phase information by PTF is relatively high overall in the whole passband. Even though the detailed cost function may not be given to evaluate the performance of transfer function, the above three parameters of PTF are enough to find the corresponding optimal illumination pattern among all combinations. Moreover, the binary coding of illumination pattern index makes it is easier to calculate the PTF and evaluate the quality of PTF.

\begin{figure}[!htp]
    \centering
    \includegraphics[width=15.5cm]{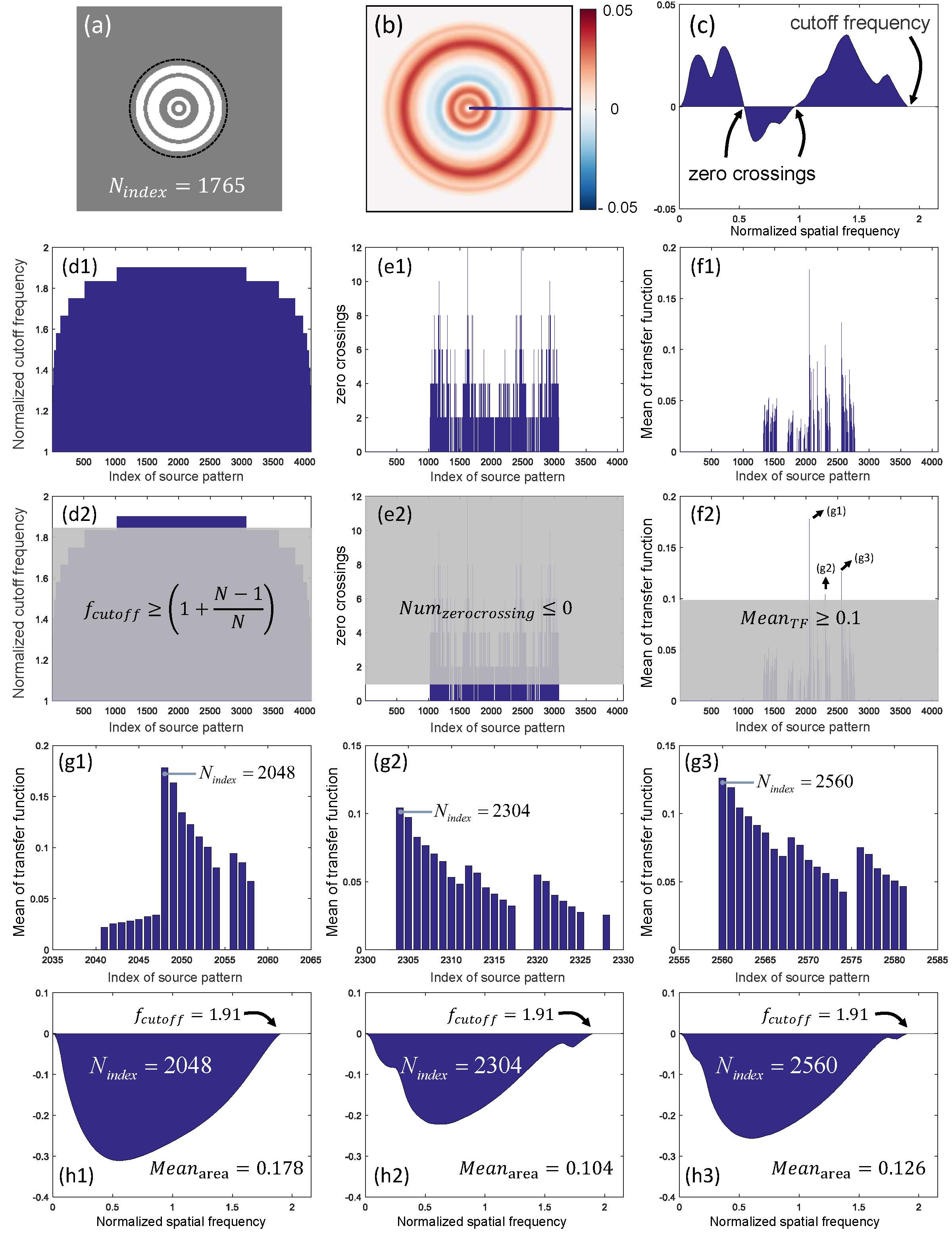}
    \caption{The imaging performance evaluation of PTF of all illumination source patterns using three crucial criteria, including cutoff frequency, zero crossings, and mean value of absolute of PTF. (a-c) Example of binary coded illumination pattern, corresponding 2D PTF image in color map, and area plot of radius cut-line of PTF. (d1-d2) The cutoff frequency value of PTF  of all illumination patterns is calculated and the illumination patterns whose cutoff frequency equals the normalized spatial threshold value ${1 + {{\left( {N - 1} \right)}/N}}$ are retained. (e1-e2) Statistics of zero crossings of source patterns with the maximum cutoff frequency value, and the preserved source patterns with no zero crossings. (f1-f2) Mean value of absolute PTF curves with both the maximum cutoff frequency and no zero crossings are calculated, and  the PTFs of top three mean values are selected. (g1-g3) Enlarged PTFs of top three mean values. (h1-h3) Area plots of radius cut-line of the top three best PTF with the same cutoff frequency value.}
    \label{Fig3}
\end{figure}

As an example, a binary coded illumination pattern and corresponding PTF image in color map are illustrated in Fig. \ref{Fig3}(a) and \ref{Fig3}(b), respectively. The area plot of radius cut-line of PTF is drawn in Fig. \ref{Fig3}(c), and the indications of zero crossings and cutoff frequency point are given in this sub-figure as well. Firstly, we calculate the PTF cutoff frequency of all illumination patterns from the index 1 to 4095 and select the PTFs whose cutoff frequency is equal to ${1 + {{\left( {N - 1} \right)} \mathord{\left/ {\vphantom {{\left( {N - 1} \right)} N}} \right. \kern-\nulldelimiterspace} N}}$. The illumination patterns with the maximum cutoff frequency value are retained, as depicted in Fig. \ref{Fig3}(d1) and \ref{Fig3}(d2). Next, the illumination pattern index mask whose cutoff frequency equals the normalized spatial threshold value ${1 + {{\left( {N - 1} \right)} \mathord{\left/ {\vphantom {{\left( {N - 1} \right)} N}} \right. \kern-\nulldelimiterspace} N}}$ is applied to the statistics of zero crossings of all source patterns, and the filtered result of zero crossings is shown in Fig. \ref{Fig3}(e1). Then, the illumination patterns with the maximum cutoff frequency value and no zero crossings are preserved in Fig. \ref{Fig3}(e2). As mentioned above, if the curve of transfer function stretch across two quadrants (first quadrant and fourth quadrant), the zero crossings make the response bad around these points because it is difficult to recover the signal around these zero points. Finally, the mean value of absolute PTF curves with both the maximum cutoff frequency value and no zero crossings are calculated as shown in Fig. \ref{Fig3}(f1). The threshold mean value of transfer function is set to be 0.1, and the PTFs of top three mean values are selected and enlarged, as illustrated in Fig. \ref{Fig3}(g1) -\ref{Fig3}(g3). It is obvious that the index of PTF with the maximum mean value is 2048, and the corresponding illumination pattern is the optimal one under the criteria for the quality of PTF. Meanwhile, the area plots of radius cut-line of the top three best PTF with the same maximum cutoff frequency and no zero crossings are illustrated as well in Fig. \ref{Fig3}(h1) -\ref{Fig3}(h3) for intuitive comparison.

\begin{figure}[!b]
    \centering
    \includegraphics[width=12.5cm]{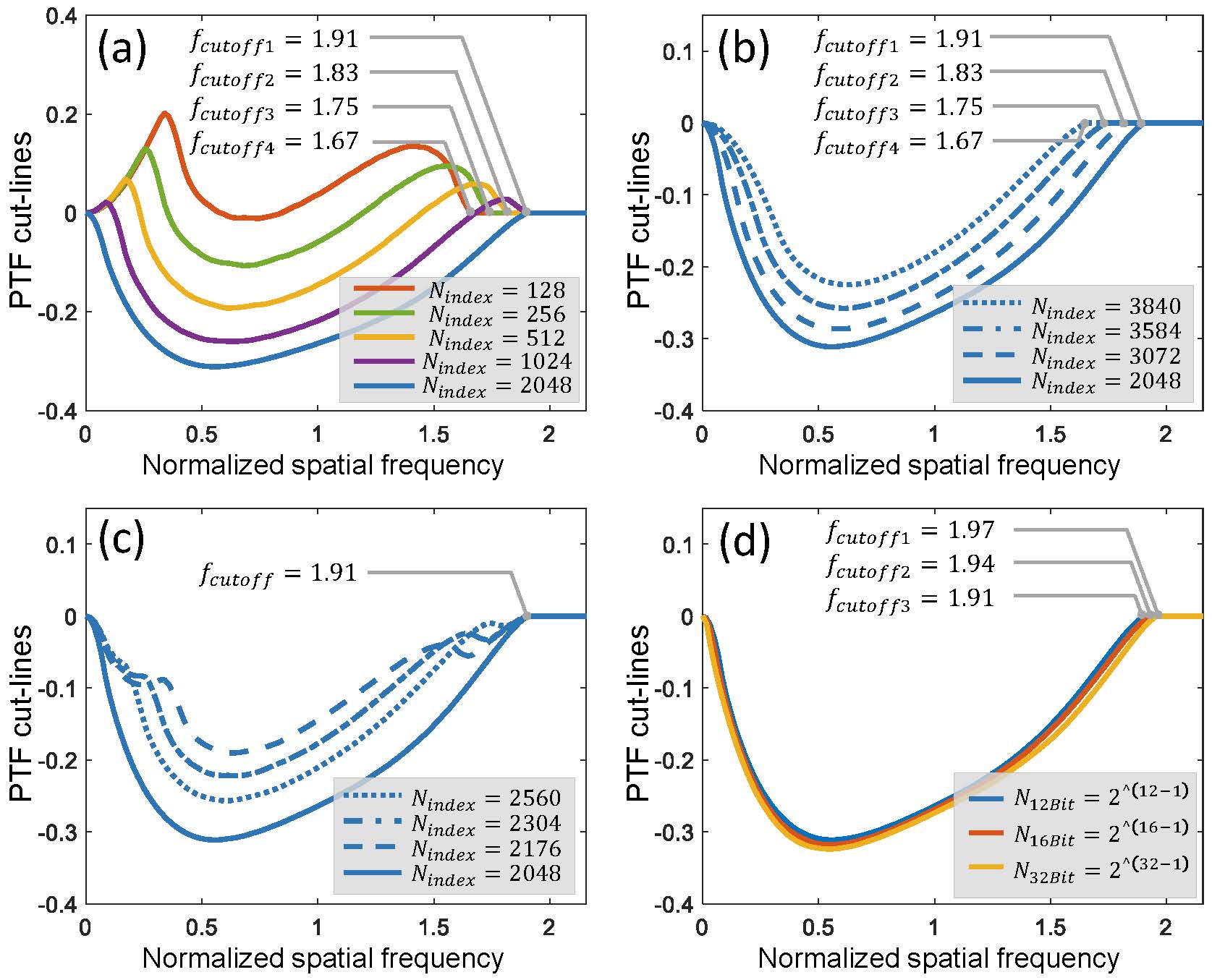}
    \caption{Comparative line profiles of PTFs of illumination patterns with different index. (a) The cut-lines of PTF of source patterns with special index from 128 (12'b0000 1000 0000) to 2048 (12'b1000 0000 0000). (b) The cut-lines of PTF of source patterns with the same maximum outer radius and different annular width from 2048 (12'b1000 0000 0000) to 3840 (12'b1111 0000 0000). (c) The cut-lines of PTF of combined source patterns from 2176 (12'b1000 1000 0000) to 2560 (12'b1010 0000 0000). (d) The cut-lines of PTF with different bit-depth (12-bit, 16-bit and 32-bit) of illumination source segmentation for single annular illumination pattern.}
    \label{Fig4}
\end{figure}

To better explain the result of the optimal illumination pattern using above three criteria of PTF's quality, we plot the cut-lines of PTF of source patterns with special index from 128 to 2048, which is composed of a single annulus with different radii in Fig. \ref{Fig4}(a). It can be seen that the curve of PTF is gradually moves from the first quadrant to the fourth quadrant, and the number of zero crossings is decreasing. Simultaneously, the cutoff frequency value of PTF increases with the increasing of source index from 128 to 2048. But the cutoff frequency of annular patterns of $N_{index} = 2048$ and $N_{index} = 1024$ are same with each other, and the illumination pattern $N_{index} = 2048$ with the biggest outer radius matched with objective NA has the best quality among these five patterns due to less zero crossings. While these single annular patterns are combined together, the combined source pattern has the maximum outer radius and different annular width, and the index of four selected illumination patterns are 2048, 3072, 3584, and 3840, respectively. Figure \ref{Fig4}(b) shows the PTFs of these illumination patterns, and the cutoff frequency and amplitude of PTF decrease with the increasing of source index from 2048 to 3840. This is because that the PTF of $N_{index} = 3072$ is the incoherent superposition of PTFs of $N_{index} = 1024$ and $N_{index} = 2048$ based on the theory of coherent mode composition. And the sign of PTF of index 1024 and 2048 are opposite with each other at both low and high frequency, so both of the response amplitude and cutoff frequency value of PTF of $N_{index} = 3072$ curve are reduced due to the superposition and offset of PTFs between $N_{index} = 2048$ and $N_{index} = 1024$. For a continuous annular illumination pattern, if the normalized outer radius of the pattern does not match with objective NA, the cutoff frequency value of PTF is determined by the outer radius of the pattern, just like the PTFs from $N_{index} = 128$ to $N_{index} = 1024$ in Fig. \ref{Fig4}(a). On the contrary, if the normalized outer radius of the illumination pattern equals objective NA, the cutoff frequency value of PTF is determined by the inner radius of the pattern, as shown in Fig. \ref{Fig4}(b). So, the cutoff frequency of PTF is decreased from 1.91 to 1.67 with the increasing of annular width of illumination pattern from $N_{index} = 2048$ to $N_{index} = 3840$ in Fig. \ref{Fig4}(b), and the magnitude of curve becomes weak gradually at both low and high frequency as well.

In Fig. \ref{Fig4}(c), the cut-lines of four representative PTFs in Fig. \ref{Fig3}(f1) are plotted, and these profiles have the same cutoff frequency value $f_{cutoff} = 1.91$ but different response amplitude. The source pattern index 2176 is the sum of index 2048 and 128, 2304 is the sum of 2048 and 256, and  2560 is the sum of 2048 and 512 respectively, so the incoherent superposition of transfer function between these PTFs leads the attenuation of response amplitude in the whole passband due to the offset contribution from the inner annular pattern. Besides, the effect of bit-depth of illumination source segmentation for single annular illumination pattern with the biggest outer radius is illustrated in Fig. \ref{Fig4}(d). The source is divided into 12-bit, 16-bit and 32-bit in binary, and the index of optimal illumination pattern is always corresponding to ${2^{N- 1}}$. These three curves are highly coincident with each other excepted that the cutoff frequency is gradually close to incoherent diffraction resolution $(2NA)$ with the increasing of bit-depth of segmentation or decreasing of annular width of illumination pattern. Moreover, the amplitude of PTF response of 32-bit at both low and high frequency is the best among these three PTFs as shown in Fig. \ref{Fig4}(d). Higher bit-depth of segmentation can provide more precise representation of arbitrary illumination pattern, but the number of source index will become larger and the calculation process of transfer function will become time-consuming. Overall, the optimal illumination pattern is the biggest single annulus matched with objective pupil and the annular width of pattern should be as small as possible. And the corresponding PTF with the best imaging performance can be available for quantitative phase reconstruction in QPI.

\subsection{Illumination pattern on LED array}

\begin{figure}[!t]
    \centering
    \includegraphics[width=15cm]{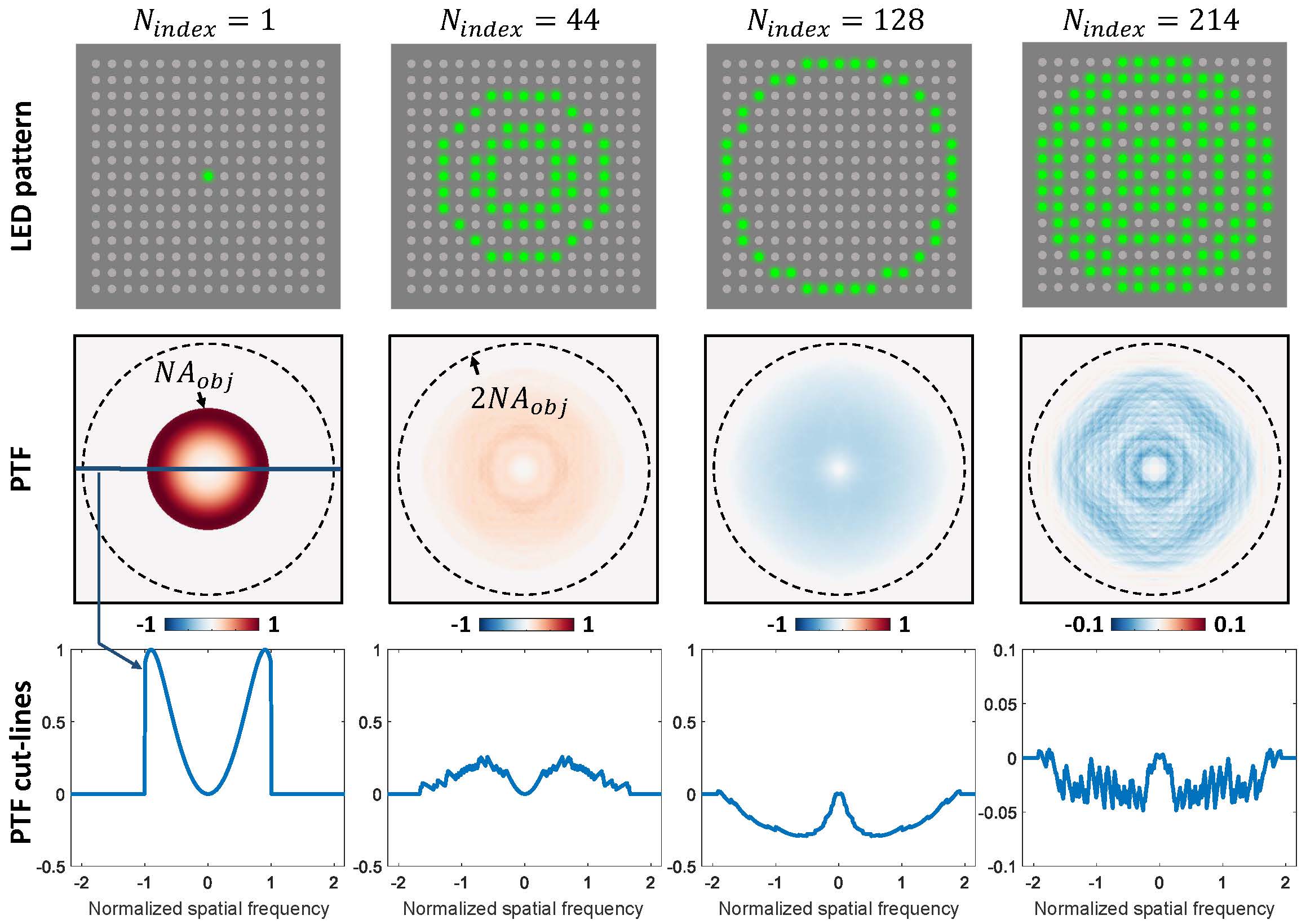}
    \caption{Four selected index of illumination pattern on the LED array. Sub-figures in the first two rows are the actual LED source pattern in light path and the corresponding discrete pattern. Last two rows of figure are the 2D images and cut lines of PTFs.}
    \label{Fig5}
\end{figure}

After the full investigation of the optimal illumination pattern in QPI and the analysis of corresponding PTF, the above mentioned binary coding method is applied to the programmable source on a LED array. Each LED on the board can be treated as a coherent source, and the light emitted from the LED is spatially coherent quasi-monochromatic illuminations approximately. And the detailed parameters of LED array used in this work will be given in the section of implementation. The size of used region of whole LED array is 15 $\times$ 15, thus the bit-depth of LED illumination pattern segmentation is 8-bit. Figure \ref{Fig5} shows four representative illumination patterns on the LED array,  and the first two rows of figure are the actual LED source patterns in light path and the corresponding discrete source patterns, respectively. And the 2D PTF images and cut-lines of PTF are plotted in the last two rows of Fig. \ref{Fig5} as well.

 The coherent situation is illustrated in the first column of figure, and the maximum response amplitude of coherent PTF is 1, but the cutoff frequency is limited by the objective pupil which corresponds to 1 in the normalized spatial frequency. As for the plots in the second and third columns of Fig. \ref{Fig5}, the curve of PTF moves from the first and second quadrants ($y > 0$) to the third and fourth quadrant ($y < 0$) in the cartesian coordinate. It's just like the relationship between the curve of $N_{index} = 128$ and the curve of $N_{index} = 2048$ in Fig. \ref{Fig4}(a) excepted that the middle frequency components of PTF of LED pattern $N_{index} = 44$ is boosted by the inner annular LEDs in Fig. \ref{Fig5}. And it is obvious that the PTF line profile of LED illumination pattern $N_{index} = 128$ are the best among these four PTFs under the above mentioned three criteria for the quality of PTF. The display range of PTF image in the last column is enlarged for better contrast, and the amplitude of curve is close to zero but oscillating in the whole passband. The source pattern of $N_{index} = 214$ is almost an incoherent illumination, and it is hard to transfer the phase information into intensity image. The incoherent superposition of each discrete LED in illumination pattern $N_{index} = 214$ on the source plane leads offset and oscillation of PTF curve. It should be noted that such oscillation of transfer function is caused by the discontinuity of isolated LED source, and the similar phenomenon is also emerged in the Fig. 5 of QPI approach in DPC \cite{DPC_LED}. Despite the relatively small bit-depth of LED illumination source segmentation, the distribution and characterization of PTF of LED source is same with the above mentioned results under the situation of more continuous light source.

\section{Implementation}
\begin{figure}[!t]
    \centering
    \includegraphics[width=16cm]{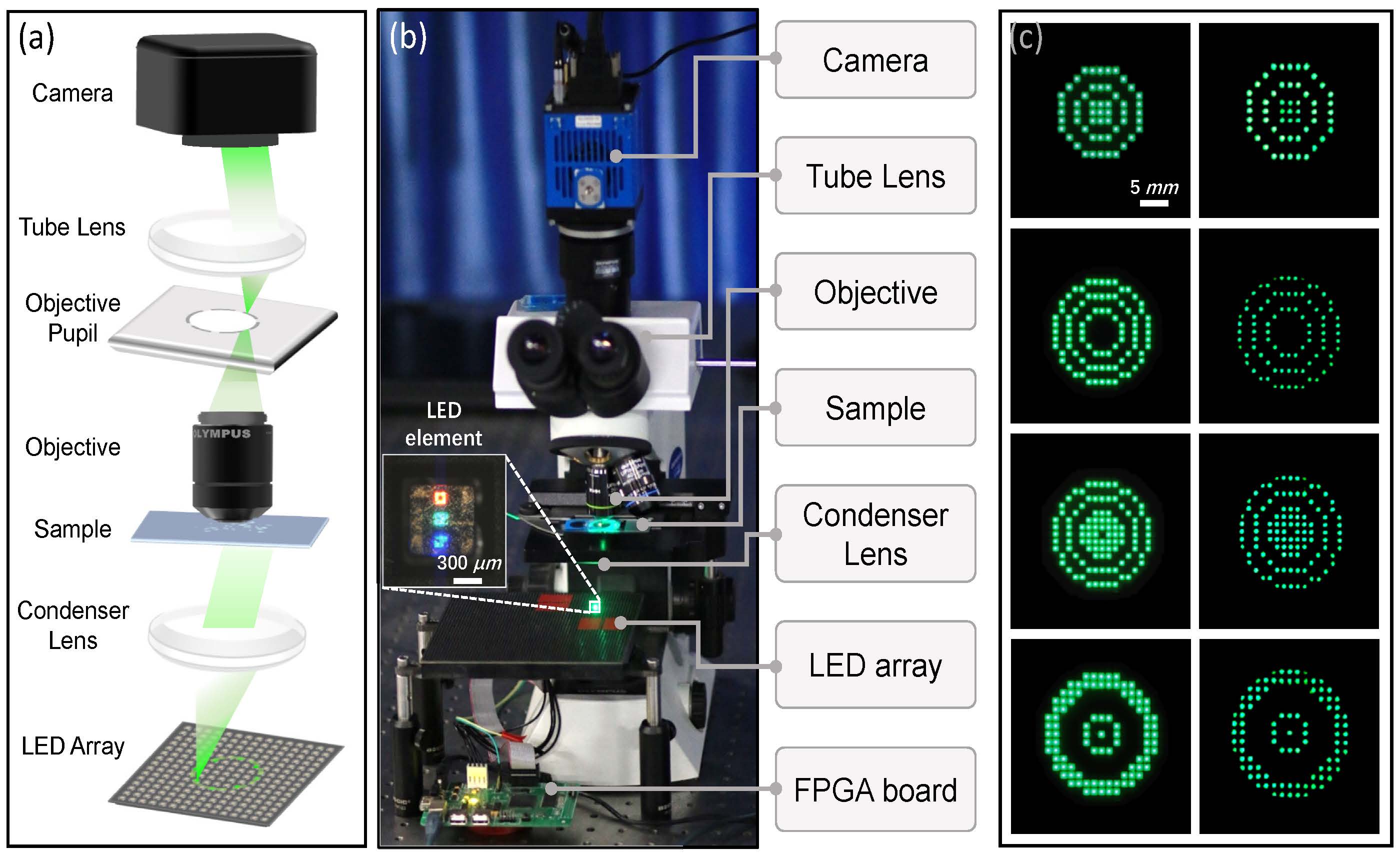}
    \caption{(a) Schematic diagram of implementation. (b) Photograph of whole imaging system. The crucial parts of setup are highlighted with the gray
    boxes. (c) Four selected binary LED illumination patterns on the LED array and the respective pattern distribution in the Fourier plane. Scale bar represents 300 $\mu$m and 10 $mm$, respectively.}
    \label{Fig6}
\end{figure}

In order to implement the proposed QPI technique based on optimal illumination pattern, we employ a traditional bright-field microscope (CX22, Olympus), and the hardware involves simply replacing the illumination unit of microscope with an LED array, as depicted in Fig. \ref{Fig6}(a). The commercial surface-mounted LED array is placed in the front focal plane of the condenser as illumination source for binary coding. Each LED can be controlled individually to illuminate the sample from a unique angle and can provide approximately spatially coherent quasi-monochromatic illuminations with narrow bandwidth (central wavelength $\lambda$ = 530 nm, $\sim$ 20 nm bandwidth). The semiconductor die of green source is located in the center position of each LED, and the size of luminous die is about 150 $\mu$m. Thus, a single LED can be approximated as a spatially coherent source, and the light emitted from the condenser lens for a single LED can be nearly treated as a plane wave. The distance between every adjacent LED elements is 1.25 mm, and the original resolution of this commercial LED array is 64 $\times$ 64. Due to the limited diameter of circular condenser aperture, only a fraction of whole array (15 $\times$ 15 LEDs) is used for binary coded illumination in experiments. Moreover, the LED array is driven dynamically by a LED controller board and this board is custom-built by ourselves with a Field Programmable Gate Array (FPGA) unit.

In this work, the microscope is equipped with an universal plan super-apochromat objective (Olympus, UPlan SAPO 20$\times$, Extra 2$\times$ magnification, NA $=$ 0.75) and a scientific CMOS (sCMOS) camera (PCO.edge 5.5, 6.5 $\mu$m pixel pitch). The photograph of whole imaging system is illustrated in Fig. \ref{Fig6}(b), and the crucial parts of setup in this photo are marked with the gray boxes. Four selected binary LED illumination patterns and respective pattern distribution in the Fourier plane are shown in Fig. \ref{Fig6}(c) as well. The discrete annular LED patterns matched with objective pupil are displayed on the LED array, and the illumination pattern is properly centered in the optical pathway. The LED patterns in the second column of Fig. \ref{Fig6}(c) are taken in the objective pupil plane by inserting a Bertrand lens into one of the eyepiece observation tubes or removing the eyepiece tubes. Moreover, a short video which contains the fast switching of binary coded LED illumination patterns captured from LED array and objective pupil is shown in \textcolor{blue}{Visualization 1}.

\section{Experimental results}
\begin{figure}[!t]
    \centering
    \includegraphics[width=15cm]{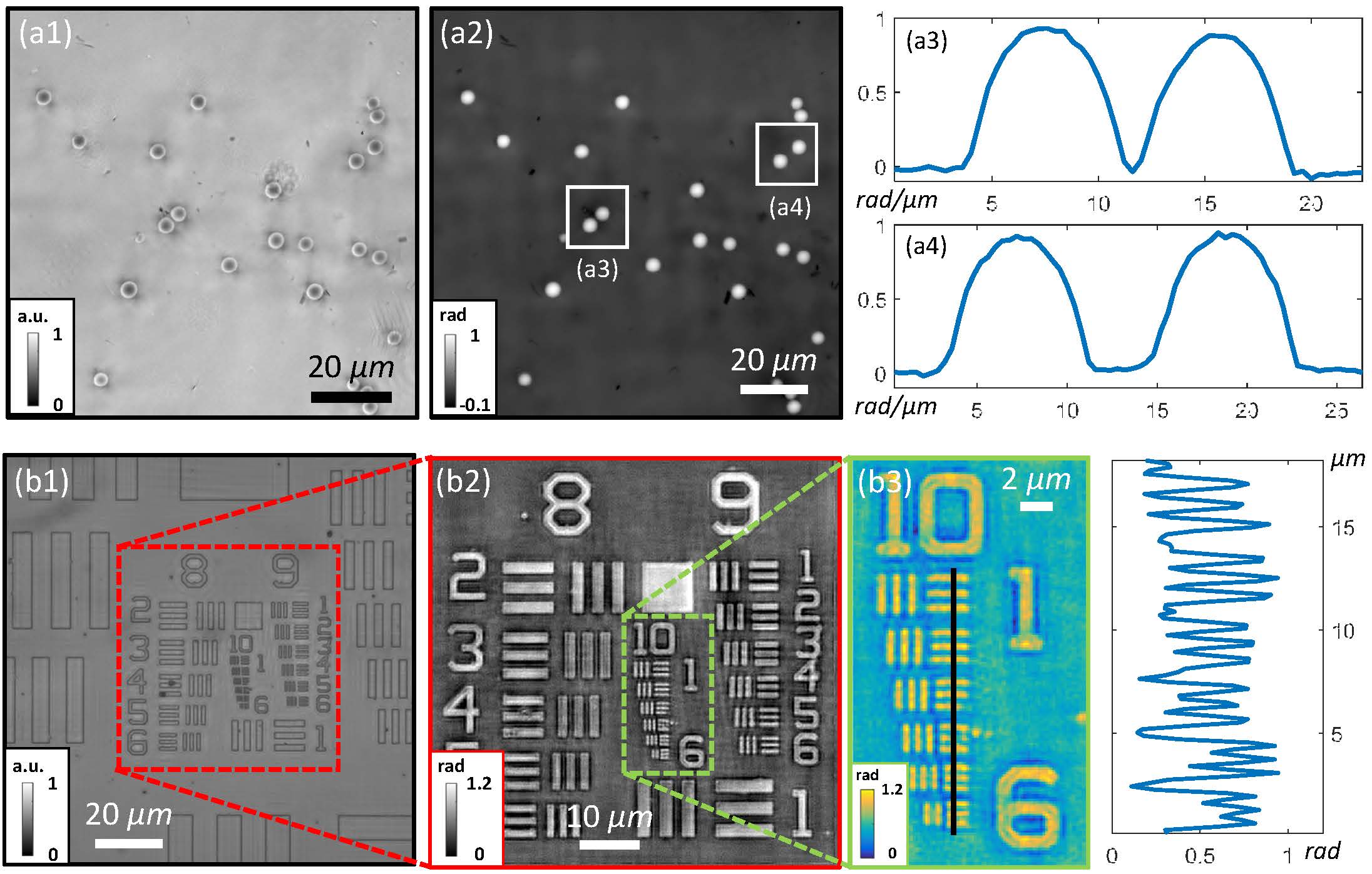}
    \caption{Quantitative phase reconstruction results based on control samples using the optimal illumination pattern ($N_{index} = 128$). (a) Results of the micro polystyrene bead with 8 $\mu$m diameter. (b) Results of quantitative phase resolution target. The achievable experimental imaging resolution is 0.274 $\mu$m, corresponding resolution element 6 in group 10 while the theoretical achievable resolution bar width is 0.188 $\mu$m, corresponding resolution element 3 in group 11. Scale bar denotes 20 $\mu$m, 10 $\mu$m, and 2 $\mu$m respectively.}
    \label{Fig7}
\end{figure}

To demonstrate the QPI capability of proposed technique based on the optimal illumination pattern, this method is applied to the quantitative phase reconstruction of micro polystyrene bead. The micro polystyrene bead (Polysciences, $n$ = 1.59) with 8 $\mu$m diameter is immersed in oil (Cargille, $n_{m}$ = 1.58), and the intensity image is measured using a 0.75 NA objective and a sCMOS camera. The sample is slightly defocused, and three intensity images are recorded at $\pm$ 0.5 $\mu$m planes and in-focus plane, accordingly. Figure \ref{Fig7}(a1) and \ref{Fig7}(a2) are the captured defocused intensity image and recovered quantitative phase image of micro polystyrene bead, respectively. For more obvious analysis results of micro phase bead, two line profiles of phase change within two close micro spheres are selected, as illustrated in Fig. \ref{Fig7}(a3) and \ref{Fig7}(a4). By comparing the two sets of line profile of phase change, the accuracy of QPI based on optimal illumination is validated.

Besides, the phase reconstruction result of quantitative phase resolution target (Benchmark Technologies Corporation, United States, Refractive Index $n = 1.52$, Height $H = 200 nm$) has also been proposed for the investigation of achievable imaging resolution using the optimal illumination pattern. Figure \ref{Fig7}(b1) is the bright-field image of full field of view (FOV) under coherent illumination, and the corresponding recovered phase image is illustrated in Fig. \ref{Fig7}(b2) as well. The resolution element group 10 is enlarged and displayed in obvious color map in Fig. \ref{Fig7}(b3), and the line profile plot of resolution elements in group 10 is illustrated as well. From the cut-line of resolution elements in group 10, we can clearly distinguish these resolution bars up to element 6. The smallest bar width of resolution element in group 10 is 0.274 $\mu$m, but the theoretical achievable bar width based under the optimal illumination pattern is 0.188 $\mu$m (1.41 effective NA, corresponding resolution element 3 in group 11). Because the smallest phase resolution bar of phase resolution target is 0.274 $\mu$m, the phase reconstruction result of this phase target is not enough for the validation of achievable resolution of proposed approach. Moreover, due to the inconspicuous improvement of imaging resolution from coherent illumination pattern (0.353 $\mu$m, corresponding group 10-3 resolution element) to optimal illumination pattern (0.274 $\mu$m, corresponding group 10-6 resolution element) on the experimental control samples (micro phase bead and quantitative phase resolution target), we do not provide the various phase reconstruction results under different illumination patterns. Even so, two sets of quantitative characterizations of control samples validate success and accuracy of our method basically.

\begin{figure}[!htp]
    \centering
    \includegraphics[width=15.5cm]{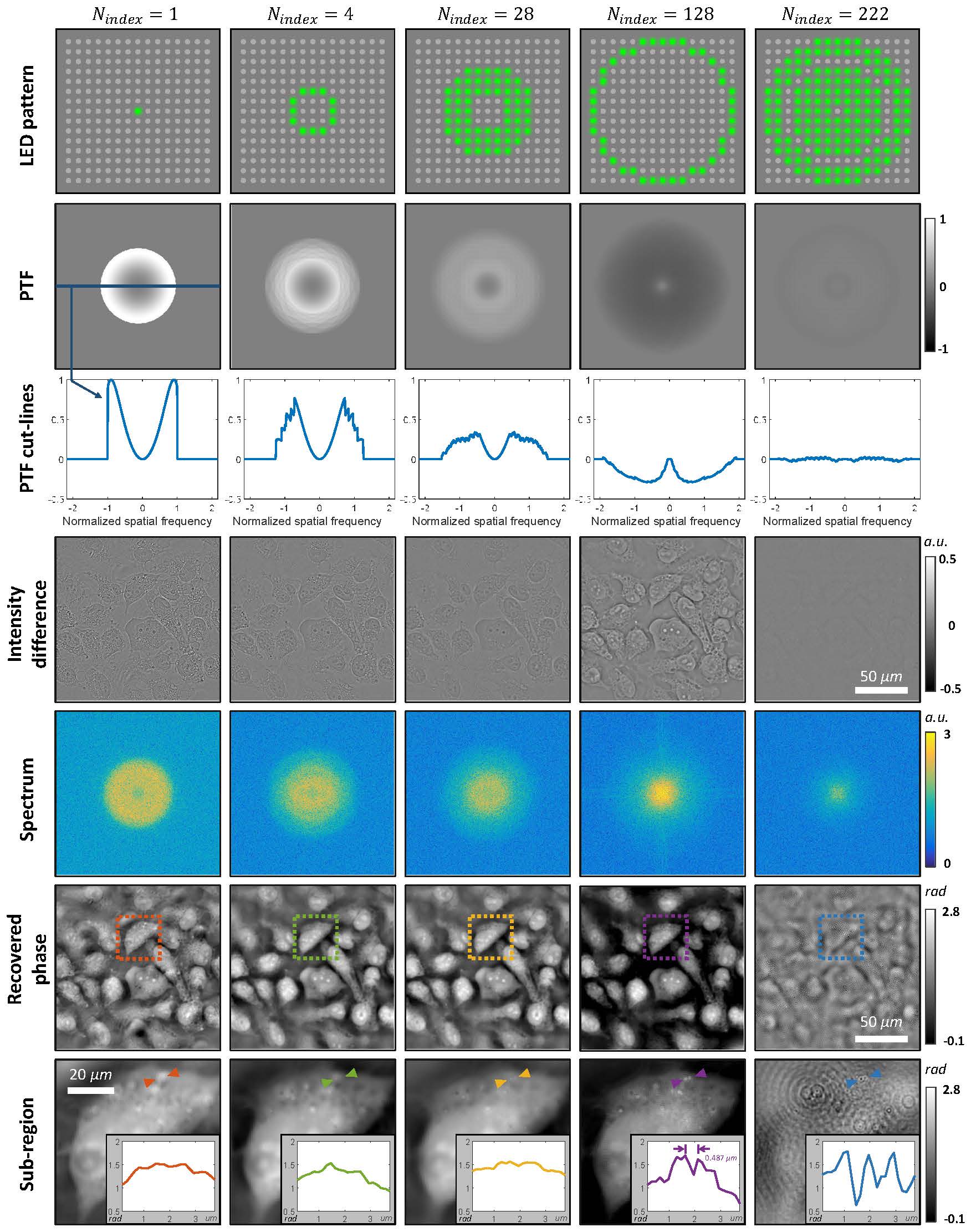}
    \caption{Experimental results of unstained HeLa cell under five selected LED illumination patterns. First three rows of sub-figures are the five selected LED illumination patterns, corresponding 2D PTF images, and cut-line profiles. Sub-figures in the fourth and fifth rows are the difference images and 2D Fourier spectrum images of HeLa cell under different illumination patterns. Last two rows of sub-figures show the recovered quantitative phase images of HeLa cell and enlarged sub-region of single cell during the interphase. Scale bar denotes 50 $\mu$m and 20 $\mu$m, respectively.}
    \label{Fig8}
\end{figure}

We also test the proposed QPI method based on the optimal illumination pattern in its intended biomedical observation experimentally. The unstained HeLa cell is imaged with different illumination patterns, and the detailed comparative results are presented in Fig. \ref{Fig8}. Five selected LED illumination patterns, corresponding 2D PTF images, as well as cut-line profiles are illustrated in the first three rows of Fig. \ref{Fig8}. Two symmetric defocused intensity images and one in-focus images are captured with 1 $\mu$m distance between two defocused intensity images. Thus, the intensity difference image of HeLa cell along $z$ axis is available by the subtraction of these two defocused intensity images, and the 2D Fourier transform is applied to intensity difference images, as depicted in the fourth and fifth rows of Fig. \ref{Fig8}. From the images of PTF and spectrum of intensity difference we can see that the shapes of spectrum and corresponding PTF are basically same with each other. This is because the frequency components in intensity difference spectrum are transferred by the PTF from phase information for a certain illumination patten. Among these five spectrum images, the low frequency amplitude of spectrum of illumination pattern with $N_{index} = 128$ is the best, and this is coincident with the robust response of corresponding PTF at low frequency.

The last two rows of Fig. \ref{Fig8} show the recovered quantitative phase images of HeLa cell under different illumination patterns, and a enlarged sub-region of single cell during the interphase. Two close tiny grains in cytoplasm are marked with arrows, and the quantitative cut lines through these grains are plotted as well. The recovered phase image under the optimal illumination pattern is better than the reconstruction results using other four illumination patterns obviously. The contrast of phase result with optimal pattern is the strongest, and this phenomenon means that the recovered low frequency components of phase object is relatively robust. Besides, the cut lines of tiny phase grains demonstrate the high frequency features in cell can still be distinguished. Although above experimental results are based on several selected illumination patterns, the comparative PTF images, intensity difference images, Fourier images and recovered phase images can prove that the annular illumination source with the biggest outer radius matched with objective pupil is the optimal illumination pattern in QPI.


\section{Conclusion and discussion}
In conclusion, we present a optimization scheme for determining the optimal illumination pattern in TIE-based QPI. In order to achieve the best phase imaging performance, the optimum illumination pattern is expected to have the maximum cutoff frequency, the minimum number of zero crossings, and the largest absolute response. By comparing the PTFs of arbitrary isotropic binary-coded illumination patterns based on the combined criterion, the optimal illumination pattern is identified, which is a thin annulus matching the NA of the objective. We implement the optimal illumination strategy on small-pitch LED array, and the QPI capability and optimality of proposed illumination pattern are demonstrated based on quantitative phase reconstructions of a phase resolution target and biological cells.

There are still several aspects that need to be addressed or require further investigations. In this work, we quantitatively assess the performance of different illumination patterns based on the responses of their PTFs. The PTFs are calculated by Eq. (\ref{Eq4}), which is normalized by the total illumination intensity (intensity of the source integrated over the pupil), thus demonstrating the "contrast" of the image (relative strengths of information-bearing portion of the image and the ever-present background). Under such conditions, we have found the optimal illumination pattern, which is the annular source with the outer radius corresponding to the objective NA, and the annulus should be as thin as possible theoretically. However, when the sample is illuminated by a low brightness LED array, using the narrower annulus means less intensity for illumination, leading to poor signal-to-noise ratio (SNR) of the captured intensity image. In order to achieve a tradeoff between the \emph{absolute intensity strength }and the \emph{normalized phase contrast}, the width of the annulus is usually designed as 5$\%$ to 10$\%$ of the radius. But it is better to consider the unnormalized PTF \cite{SheppardInterpretation}, which can give the absolute value of the image signal and determine its strength relative to the noise level (the annular illumination may no longer be the optimal solution). We leave this interesting avenue for future investigation.


\section*{ACKNOWLEDGMENTS}
This work was supported by the National Natural Science Fund of China (61722506, 61505081, 111574152), Final Assembly `13th Five-Year Plan' Advanced Research Project of China (30102070102), National Defense Science and Technology Foundation of China (0106173), Outstanding Youth Foundation of Jiangsu Province of China (BK20170034), The Key Research and Development Program of Jiangsu Province, China (BE2017162), `Six Talent Peaks' project of Jiangsu Province, China (2015-DZXX-009), `333 Engineering' Research Project of Jiangsu Province, China (BRA2016407), Fundamental Research Funds for the Central Universities (30917011204, 30916011322), Open Research Fund of Jiangsu Key Laboratory of Spectral Imaging \& Intelligent Sense (3091601410414).

\section*{DISCLOSURES}
The authors declare that there are no conflicts of interest related to this article.

\end{document}